\documentclass[aps,prl,twocolumn,superscriptaddress, showpacs]{revtex4}

\usepackage{graphicx}

\begin{document}

\title{Nature of magnetic coupling between Mn ions in as-grown Ga$_{1-x}$Mn$_{x}$As studied by x-ray magnetic circular dichroism}

\author{Y. Takeda}
\email[]{ytakeda@spring8.or.jp}
\affiliation{Japan Atomic Energy Agency, Synchrotron Radiation Research Center SPring-8,  Mikazuki, Hyogo 679-5148, Japan}
\author{M. Kobayashi}
\affiliation{Department of Physics, The University of Tokyo, ,Hongo, Tokyo 113-0033, Japan}
\author{T. Okane}
\affiliation{Japan Atomic Energy Agency, Synchrotron Radiation Research Center SPring-8,  Mikazuki, Hyogo 679-5148, Japan}
\author{T. Ohkochi}
\affiliation{Japan Atomic Energy Agency, Synchrotron Radiation Research Center SPring-8,  Mikazuki, Hyogo 679-5148, Japan}
\author{J. Okamoto}
\altaffiliation{National Synchrotron Radiation Research Center, 101 Hsin-Ann Road, Hsinshuu Science Park, Hsinchu 30077, Taiwan, R. O. C.}
\affiliation{Japan Atomic Energy Agency, Synchrotron Radiation Research Center SPring-8,  Mikazuki, Hyogo 679-5148, Japan}
\author{Y. Saitoh}
\affiliation{Japan Atomic Energy Agency, Synchrotron Radiation Research Center SPring-8,  Mikazuki, Hyogo 679-5148, Japan}
\author{K. Kobayashi}
\affiliation{Japan Atomic Energy Agency, Synchrotron Radiation Research Center SPring-8,  Mikazuki, Hyogo 679-5148, Japan}
\author{H. Yamagami}
\affiliation{Japan Atomic Energy Agency, Synchrotron Radiation Research Center SPring-8,  Mikazuki, Hyogo 679-5148, Japan}
\author{A. Fujimori}
\affiliation{Department of Physics, The University of Tokyo, ,Hongo, Tokyo 113-0033, Japan}
\author{A. Tanaka}
\affiliation{Graduate School of Advanced Sciences of Matter, Hiroshima University, Higashi-Hiroshima 739-8530, Japan}
\author{J. Okabayashi}
\altaffiliation{Department of Physics, Tokyo Institute of Technology, Ookayama, Meguro-ku, Tokyo 152-8551, Japan}
\affiliation{Department of Applied Chemistry, The University of Tokyo, Hongo, Tokyo 113-8656, Japan}
\author{M. Oshima}
\affiliation{Department of Applied Chemistry, The University of Tokyo, Hongo, Tokyo 113-8656, Japan}
\author{S. Ohya}
\affiliation{Department of Electronic Engineering, The University of Tokyo, Hongo, Tokyo 113-8656, Japan}
\affiliation{PRESTO JST, Kawaguchi, Saitama 331-0012, Japan}
\author{P. N. Hai}
\affiliation{Department of Electronic Engineering, The University of Tokyo, Hongo, Tokyo 113-8656, Japan}
\author{M. Tanaka}
\affiliation{Department of Electronic Engineering, The University of Tokyo, Hongo, Tokyo 113-8656, Japan}

\date{\today}

\begin{abstract}
The magnetic properties of as-grown Ga$_{1-x}$Mn$_{x}$As have been investigated by the systematic measurements of temperature and magnetic field dependent soft x-ray magnetic circular dichroism (XMCD). 
The {\it intrinsic} XMCD intensity at high temperatures obeys the Curie-Weiss law, but residual spin magnetic moment appears already around 100 K, significantly above Curie temperature ($T_C$), suggesting that short-range ferromagnetic correlations are developed above $T_C$. 
The present results also suggest that antiferromagnetic interaction between the substitutional and interstitial Mn (Mn$_{int}$) ions exists and that the amount of the Mn$_{int}$ affects $T_C$.
\end{abstract}


\maketitle

Ga$_{1-x}$Mn$_{x}$As is a prototypical and most well-characterized diluted magnetic semiconductor (DMS) \cite{Ohno_Science_1998}. 
Because Ga$_{1-x}$Mn$_{x}$As is grown under thermal non-equilibrium conditions, however, it is difficult to avoid the formation of various kinds of defects and/or disorder. 
In fact, Rutherford backscattering (RBS) channeling experiments for as-grown Ga$_{0.92}$Mn$_{0.08}$As samples has shown that as many as $\sim17$ ${\%}$ of the total Mn ions may occupy the interstitial sites \cite{Yu_PRB_2002}. 
It is therefore supposed that antiferromagnetic (AF) interaction between the substitutional Mn (Mn$_{sub}$) ions and interstitial Mn (Mn$_{int}$) ions may suppress the magnetic moment \cite{Blinowski_PRB_2003, Masek_PRB_2004}.
In addition, the random substitution of Mn ions may create inhomogeneous Mn density distribution, which may lead to the development of ferromagnetic domains above Curie temperature ($T_C$) \cite{Mayr_PRB_2002}. 
The characterization of non-ferromagnetic Mn ions is therefore a clue to identify how they are related with the ferromagnetic ordering and eventually to improve the ferromagnetic properties of Ga$_{1-x}$Mn$_{x}$As samples. 
However, it has been difficult to extract the above information through conventional magnetization measurement due to the large diamagnetic response of the substrate and the unavoidable mixture of magnetic impurities.

X-ray magnetic circular dichroism (XMCD), which is an element specific magnetic probe, is a powerful technique to address the above issues. 
So far, several results of XMCD measurements on Ga$_{1-x}$Mn$_{x}$As have been reported \cite{Ohldag_APL_2000, Edmonds_APL_2004, Edmonds_PRB_2005}. 
From $H$ dependent XMCD studies, the enhancement of XMCD intensity by post-annealing implies AF interaction between the Mn$_{sub}$ and Mn$_{int}$ ions \cite{Edmonds_PRB_2005}. 
In the present study, in order to characterize the magnetic behaviors of the Mn$_{sub}$ and Mn$_{int}$, we have extended the approach and performed systematic temperature ($T$) and magnetic field ($H$) dependent XMCD studies in the Mn $L_{2,3}$ absorption edge region of Ga$_{1-x}$Mn$_{x}$As. 
We have found that short-range ferromagnetic correlations develop significantly above $T_C$ and that AF interaction between the Mn$_{sub}$ and Mn$_{int}$ is important to understand the magnetic properties of Ga$_{1-x}$Mn$_{x}$As. 

We prepared two as-grown samples with different Mn concentrations; $x=$ 0.042 and 0.078, whose $T_C$ was $\sim$ 60 and 40 K, respectively, as determined by an Arrott plot of the anomalous Hall effect. 
To avoid surface oxidation, the sample had been covered immediately after the growth of Ga$_{1-x}$Mn$_{x}$As films by cap layers without exposure to air [As cap/GaAs cap (1nm)/ Ga$_{1-x}$Mn$_{x}$As(20nm)/GaAs(001)]. 
The X-ray absorption spectroscopy (XAS) and XMCD measurements were performed at the helical undulator beam line BL23SU of SPring-8 \cite{Yokoya_JSR_1998}. 
The XAS spectra were obtained by the total-electron yield mode. 
The measurements were done without a surface treatment and $H$ was applied to the sample perpendicular to the film surface. 

\begin{figure}
\includegraphics[scale=0.38]{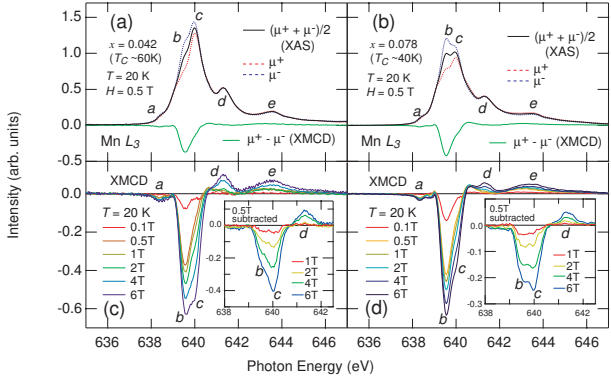}
\caption{\label{fig:Fig1} (Color online) Mn $L_3$-edge XAS ($\mu^{+}$, $\mu^{-}$ and ($\mu^{+}$ + $\mu^{-}$)/2) and XMCD ($\mu^{+}$ - $\mu^{-}$) spectra of Ga$_{1-x}$Mn$_{x}$As taken at $T=$ 20 K and $H=$ 0.5 T for $x=$ 0.042 (a) and $x=$ 0.078 (b). Panels (c) and (d) show the $H$ dependence of the XMCD spectra for $x=$ 0.042 and $x=$ 0.078, respectively. Inset shows the difference XMCD spectra obtained by subtracting the XMCD spectrum at $H=$ 0.5 T.}
\end{figure}

Figures 1 (a) and 1 (b) show the XAS spectra ($\mu^{+}$ and $\mu^{-}$) in the photon energy region of the Mn $L_3$ absorption edge and the corresponding XMCD spectra, defined as $\mu^{+}-\mu^{-}$, at $T=$ 20 K and $H=$0.5 T for $x=$ 0.042 and 0.078. 
Here, $\mu^{+}$ ($\mu^{-}$) refers to the absorption coefficient for the photon helicity parallel (anti-parallel) to the Mn 3$d$ majority spin direction. 
The XAS spectra for both Mn concentrations have five structures labeled as {\it a, b, c, d} and {\it e}. 
The average XAS spectra [defined by $(\mu^{+}$ + $\mu^{-})/2$] have been normalized to 1 at structure {\it b}. 
The intensity ratio {\it c}/{\it b} is very different between $x=$ 0.042 and 0.078, indicating that the spectra consist of two overlapping components. 
Figures 1 (c) and 1 (d) show the $H$ dependence of the XMCD spectra. 
As $H$ increases, XMCD structures corresponding to structures {\it c, d} and {\it e} are enhanced, particularly, strongly for the $x=$ 0.042 sample.
One can see this behavior more clearly in the difference XMCD spectra obtained by subtracting the XMCD spectrum at 0.5 T from the spectra at $H=$ 1, 2, 4 and 6 T as shown in the inset of Fig.1 (c) and (d). 
Recent XAS and XMCD studies have revealed that these structures ({\it c, d, e}) are ascribed to contamination of out-diffused Mn ions on the surface \cite{Edmonds_APL_2004, Ishiwata_PRB_2005, Wu_PRB_2005}. 
The difference in the XAS intensity ratio {\it c}/{\it b} is therefore naturally ascribed to the difference in the amount of Mn ions diffused into the cap layer or the surface region during the growth of GaAs on Ga$_{1-x}$Mn$_{x}$As. 
In the following, therefore, we shall neglect those extrinsic signals and focus only on intrinsic signals, particularly structure {\it b}, to investigate the intrinsic magnetic behavior. 

In order to extract the {\it intrinsic} XAS spectrum, we assumed that structure {\it b} could be ascribed to the intrinsic Mn ions as mentioned above. 
Therefore, we first obtained the {\it extrinsic} XAS spectrum as (XAS $x=$ 0.042)$-p\times$(XAS $x=$ 0.078), where $p$ was chosen so that structure {\it b} vanished. 
The {\it intrinsic} XAS spectrum was then obtained as ({\it raw} XAS)$-q\times$({\it extrinsic} XAS), where $q$ was determined so that the line shape of the {\it intrinsic} XAS spectrum agreed with that obtained from the fluorescence yield measurements \cite{Ishiwata_PRB_2005, Wu_PRB_2005}. 
Next, in order to extract the {\it intrinsic} XMCD spectra, we first obtained the {\it extrinsic} XMCD spectrum as (XMCD at 6 T)$-\alpha\times$(XMCD at 0.5 T), where $\alpha$ was chosen so that an XMCD structure corresponding to structure {\it b} vanished by utilizing the fact that the ferromagnetic signals and hence the intrinsic signals should be dominant in the XMCD spectrum at low $H$. 
The {\it intrinsic} XMCD spectrum was then obtained as (XMCD at each $H$)$-\beta\times$({\it extrinsic} XMCD spectrum), where $\beta$ was chosen so that structure {\it c} vanished. 
Figures 2 (a) and 2 (b) show the results of the decomposition of the XAS and XMCD spectra into the intrinsic and extrinsic components for $x=$ 0.042 and 0.078, respectively. 
While the XMCD intensity is enhanced as $H$ increases and $T$ decreases, the line shapes of the {\it intrinsic} XMCD spectra are unchanged with $H$ and $T$. 
The line shapes of the {\it intrinsic} XAS and XMCD spectra for both Mn concentrations thus agree with each other as shown in Fig. 2 (c), indicating that the decomposition procedure was valid. 

\begin{figure}
\includegraphics[scale=0.35]{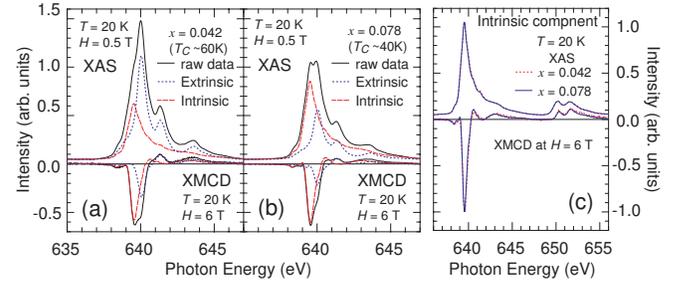}
\caption{\label{fig:Fig2} (Color online) Decomposition of the XAS and XMCD spectra of Ga$_{1-x}$Mn$_{x}$As into the intrinsic and extrinsic components for $x=$ 0.042 (a) and for $x=$ 0.078 (b) in the Mn $L_3$ edge region. Panel (c) shows comparison of the line shapes of the {\it intrinsic} XAS and XMCD spectra between $x=$ 0.042 and 0.078, normalized to the peak heights.}
\end{figure}

Using the {\it intrinsic} XAS and XMCD spectra, we have applied the XMCD sum rules \cite{Thole_PRL_1992, Carra_PRL_1993}, assuming the Mn 3$d$ electron number $N_{d}=5.1$ \cite{Edmonds_PRB_2005}, and estimated the spin magnetic moment ($M_{S}$) at $T=$ 20 K and $H=$ 0.5 T to be $M_{S}$=2.5$\pm$0.2 and 1.7$\pm$0.2 ($\mu_B$ per Mn) for $x=$ 0.042 and 0.078, respectively. 
These $M_{S}$ values are much larger than those obtained in the early studies on oxidized surfaces \cite{Ohldag_APL_2000} and comparable to the recent ones on etched surfaces \cite{Edmonds_PRB_2005}, indicating that the cap layer protected the ferromagnetic properties of Ga$_{1-x}$Mn$_{x}$As.  
The ratio $M_{L}$/$M_{S}$ is estimated to be 0.07 for both concentrations, where $M_{L}$ is the value of the orbital magnetic moment, showing that the intrinsic Mn ion has a finite, although small, $M_{L}$, probably because of certain deviation from the pure Mn$^{2+}$ ($d^5$) state. 

\begin{figure}
\includegraphics[scale=0.4]{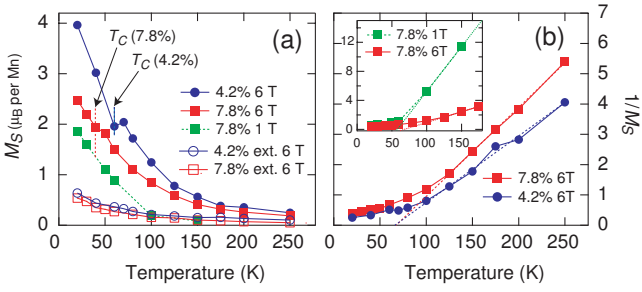}
\caption{\label{fig:Fig3} (Color online) $T$ dependence of the spin magnetic moment $M_{S}$. (a) $T$ dependence of $M_{S}$ for $H=$ 6 T. For $x=$ 0.078, results for $H=$ 1 T are also plotted. Open symbols show that of the extrinsic component at $H=$ 6 T. (b) $T$ dependence of the inverse of $M_{S}$. Inset shows comparison between 1 and 6 T for $x=$ 0.078.}
\end{figure}

The $T$ dependence of $M_{S}$ from the XMCD signal for $H=$ 6 T is plotted in Fig. 3 (a). 
As $T$ decreases, the XMCD signal is increased monotonously except for the discontinuity at around $T_C$ ($\sim60$ K for $x=$ 0.042, $\sim40$ K for $x=$ 0.078). 
This discontinuity probably reflects the ferromagnetic ordering which aligns the magnetization parallel to the sample surface, the easy axis of magnetization in the films \cite{Ohno_APL_1996}. 
It should be noted that $M_{S}$ increases monotonously even well below $T_C$ as $T$ decreases, indicating that full spin polarization is not achieved even well below $T_C$. 
For $x=$ 0.078, the $T$ dependence for $H=$ 1 T shows essentially the same behavior as that for 6 T.
Figure 3 (b) shows the inverse of $M_{S}$ plotted in Fig 3 (a). 
The high-temperature part is well described by the Curie-Weiss (CW) law, independent of $H$ as shown in the inset of Fig. 3 (b). 

Figure 4 shows the $H$ dependence of $M_{S}$ at several temperatures for $x=$ 0.042 [panel (a)] and 0.078 [panel (b)]. 
$M_{S}$ of the intrinsic component is increased rapidly from $H=$ 0.1 to 0.5 T, due to the re-orientation of the ferromagnetic moment from the in-plane to out-of-plane directions \cite{Ohno_APL_1996}. 
Above 0.5 T, $M_{S}$ is increased almost linearly as a function of $H$.
We have plotted the $T$ dependence of $M_{S}$$|_{H \rightarrow 0 \mathrm{T}}$ obtained from the linear extrapolation of $M_{S}$ at high fields to $H=$ 0 T and ${\partial}M_{S}/{\partial}H|_{H > 0.5 \mathrm{T}}$ ($\mu_B$/T per Mn) (the susceptibility of the paramagnetic component) in Fig. 4 (c) and (d), respectively. 
For the extrinsic component, $M_{S}$$|_{H \rightarrow 0 \mathrm{T}}$ is vanishingly small at all temperatures and ${\partial}M_{S}/{\partial}H|_{H > 0.5 \mathrm{T}}$ is increased as $T$ decreases following the CW law, indicating that the extrinsic component is paramagnetic and decoupled from the ferromagnetism of the intrinsic component. 
As for the ferromagnetic component, $M_{S}$$|_{H \rightarrow 0 \mathrm{T}}$ is steeply increased below $\sim100$ K. i.e., from somewhat above $T_C$. 
The $T$ dependence of $M_{S}$$|_{H \rightarrow 0 \mathrm{T}}$ [Fig. 4 (c)] is correlated with the deviation from the CW law below $\sim100$ K [Fig. 3 (b)]. 
Well below $T_C$, $M_{S}$$|_{H \rightarrow 0 \mathrm{T}}$ still continues to increase with decreasing $T$, indicating the inhomogeneous nature of the ferromagnetism. 
As for ${\partial}M_{S}/{\partial}H|_{H > 0.5 \mathrm{T}}$, unlike the extrinsic component, it saturates around $T_C$ and is not increased as $T$ further decreases.
The appearance and increase of $M_{S}$$|_{H \rightarrow 0 \mathrm{T}}$ between $T_C$ and $\sim$100 K [Fig. 4 (c)] strongly suggest that short-range ferromagnetic correlations start to develop and ferromagnetic domains form before the long-range order is established at $\it{macroscopic}$ $T_C$. 
Each ferromagnetic domain may have different ferromagnetic behavior due to the spatial distribution of $T_C$ in the as-grown samples. 
Those results may correspond to the theoretical prediction that ferromagnetic domains develop above $T_C$ when there is magnetic inhomogeneity \cite{Mayr_PRB_2002}. 

\begin{figure}
\includegraphics[scale=0.32]{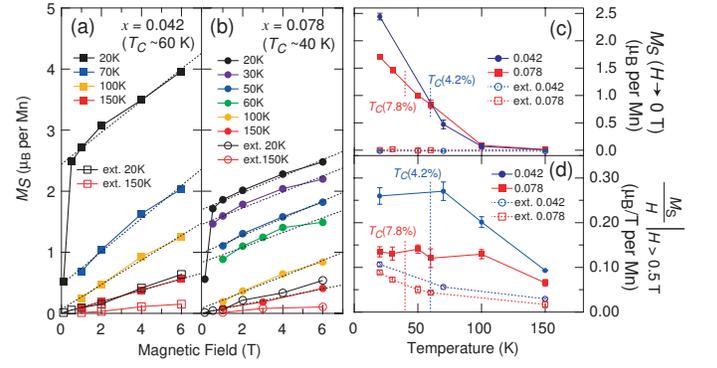}
\caption{\label{fig:Fig4} (Color online) $H$ dependence of $M_{S}$ for $x=$ 0.042 (a) and for $x=$ 0.078 (b) at several temperatures. Dashed lines show fitted straight lines above 0.5 T. (c) $T$ dependence of the residual magnetization $M_{S}$$|_{H \rightarrow 0 \mathrm{T}}$ ($M_{S}$ for $H$ $\rightarrow$ 0 T). Open symbols show that of the extrinsic component. (d) $T$ dependence of the slope of the $M_{S}$-$H$ curve above 0.5 T, i.e., the high-field magnetic susceptibility (${\partial}M_{S}/{\partial}H$$|_{H > 0.5 \mathrm{T}}$). Open symbols show that of the extrinsic component.}
\end{figure}

The suppression of the CW-like increase of ${\partial}M_{S}/{\partial}H|_{H > 0.5 \mathrm{T}}$ below $T_C$ in both samples indicates that AF interaction between the ferromagnetic Mn i.e., Mn$_{sub}$ and non-ferromagnetic (or paramagnetic) Mn such as Mn$_{int}$. 
The recent $H$ dependent XMCD study of Ga$_{1-x}$Mn$_{x}$As shows that ${\partial}M_{S}/{\partial}H|_{H > 0.5 \mathrm{T}}$ becomes small and $M_{S}$$|_{H \rightarrow 0 \mathrm{T}}$ becomes large after post-annealing, suggesting that the changes are caused by a reduction of Mn$_{int}$ \cite{Edmonds_PRB_2005}. 
In the present study, ${\partial}M_{S}/{\partial}H|_{H > 0.5 \mathrm{T}}$ and $M_{S}$$|_{H \rightarrow 0 \mathrm{T}}$ are smaller for $x=$ 0.078 than for $x=$ 0.042 [Fig. 4 (c) and (d)], suggesting that AF interaction becomes stronger for $x=$ 0.078 than that for $x=$ 0.042. 
This is reasonable because the number of Mn$_{int}$ is expected to be larger for larger Mn concentration. 
Assuming that $M_{S}$ per the Mn$_{sub}$ is 5 ($\mu_B$ per Mn) and $M_{S}$ of the Mn$_{int}$ is antiparalleled to that of Mn$_{sub}$, the ratio of Mn$_{int}$ atoms in the intrinsic component ($R_{int}$) is estimated as 0.26 for $x=$ 0.042 and 0.33 for $x=$ 0.078 from $M_{S}$$|_{H \rightarrow 0 \mathrm{T}}$ at 20 K.
This is consistent with the result of the RBS experiment \cite{Yu_PRB_2002}, which $R_{int}$ is estimated as 0.17 for an as-grown sample with $T_C$= 67 K, indicating that $T_C$ is strongly correlated with the amount of Mn$_{int}$. 
We have fitted the susceptibility ${\partial}M_{S}/{\partial}H|_{H = 6 \mathrm{T}}$ ($\mu_B$/T per Mn) of the intrinsic component above 100 K [Fig. 3 (b)] to the CW law with an offset, ${\partial}M_{S}/{\partial}H|_{H = 6 \mathrm{T}}=N_{x}C/(T-\Theta)+{\partial}M_{S}/{\partial}H|_0$, where $C=(g\mu_{B})^{2}S(S+1)/3k_{B}$ is the Curie constant, $\Theta$ is the Weiss temperature, ${\partial}M_{S}/{\partial}H|_{0}$ is the constant offset, $N_{x}$ is the number of magnetic Mn ions in the sample with Mn concentration $x$, and $g$ is the $g$ factor. 
$\Theta$ is estimated to be $68{\pm}5$ K for $x=$ 0.042 and $69{\pm}3$ K for $x=$ 0.078. 
${\partial}M_{S}/{\partial}H|_{0}$ is estimated to be of order of ${\sim}10^{-3}$ for both samples.
Assuming $g=$ 2, $S=$ 5/2 and $\Theta=$ 68 K, one obtains $N_{0.042}=$ 0.97 and $N_{0.078}=$ 0.67. 
This result strongly suggests that most of the intrinsic Mn ions in the $x=$ 0.042 sample participate in the paramagnetism above $\sim$100 K and the paramagnetism in the $x=$ 0.078 sample is suppressed even at high temperatures, again implying that the AF interaction is stronger and more influential in the $x=$0.078 sample. 

In conclusion, we have investigated the $T$, $H$ and Mn concentration dependences of the ferromagnetism in as-grown Ga$_{1-x}$Mn$_{x}$As samples by XMCD measurements to extract the intrinsic magnetic component. 
The XMCD intensity deviates from the CW law below $\sim$100 K, indicating that the ferromagnetic moment starts to form at $\sim$100K and that the short-range ferromagnetic correlations develop significantly above $T_C$. 
The high-field magnetic susceptibility becomes $T$-independent below $T_C$, indicating that the AF interaction between the Mn$_{sub}$ and Mn$_{int}$ ions, which becomes strong as the Mn concentration $x$ increases, plays an important role to determine the magnetic behavior of Ga$_{1-x}$Mn$_{x}$As. 
In addition, the amount of the Mn$_{int}$ ions should be strongly related with $T_C$.
The present experimental findings should give valuable insight into the inhomogeneous magnetic properties of many DMS's.
In future studies, it is very important to perform a detail $T$ and $H$ dependent XMCD study for a post-annealed samples.

This work was supported partly by Grants-in-Aids for Scientific Research in Priority Area ``Semiconductor Spintronics'' (14076209), ``Creation and Control of Spin Current'' (190481012) and by PRESTO/SORST of JST from the Ministry of Education, Culture, Sports, Science and Technology.

\end{document}